\documentclass[conference]{IEEEtran}
\usepackage{amssymb}
\usepackage{amsmath}
\usepackage{amsfonts}
\usepackage{graphicx}
\usepackage{algorithm}
\usepackage{algorithmic}
\usepackage{epstopdf}
\usepackage{color}
\usepackage{multirow}
\usepackage{lettrine}
\usepackage{caption}
\usepackage{subcaption}
\usepackage{amsmath,amsfonts,amssymb,amsthm, bm}
\usepackage{xcolor,soul,framed}
\usepackage[noadjust]{cite}
\usepackage{filecontents}
\usepackage{multibib}
\usepackage{relsize}
\normalsize
\usepackage{multicol,lipsum,xparse}

\ifCLASSINFOpdf
\else
\fi
\begin{document}
%
\title{SCMA Spectral and Energy Efficiency with QoS}

\author{
	Samira~Jaber, 
        and Wen~Chen, 
\\
        Shanghai Institute of Advanced Communications and Data Sciences\\
Department of Elecronic Engineering, Shanghai Jiao Tong University, China\\
Email: \{samira; wenchen\}{\@}sjtu.edu.cn\\
Kunlun Wang\\
Shool of Information Sciences and Technology, ShanghaiTech University, China\\
Email: wangkl2{\@}shanghaitech.edu.cn\\
Qingqing Wu\\
Department of Electrical and Computer Engineering, University of Macau, China\\
Email: qingqingwu@um.edu.mo
        
        \thanks{S. Jaber and W. Chen are with the Department of Electronic Engineering, Shanghai Jiao Tong University, (email: \{samira, wenchen\}@sjtu.edu.cn).}
       \thanks{K. Wang is with the School of Information Science and Technology, ShanghaiTech Unitersity, (e-mails: wangkl2@shanghaitech.edu.cn).}
       \thanks{Q. Wu is with the Department of Electrical and Computer Engineering, National University of Singapore (e-mail: elewuqq@nus.edu.sg).}       
    }
%
\markboth{***}%
{Submitted paper}
\maketitle
\begin{abstract}
Sparse code multiple access (SCMA) is one of the promising candidates for new radio access interface. The new generation communication system is expected to support massive user access with high capacity. However, there are numerous problems and barriers to achieve optimal performance, e.g., the multiuser interference and high power consumption. In this paper, we present optimization methods to enhance the spectral and energy efficiency for SCMA with individual rate requirements. 
The proposed method has shown a better network mapping matrix based on power allocation and codebook assignment. Moreover, the proposed method is compared with orthogonal frequency division multiple access (OFDMA) and code division multiple access (CDMA) in terms of spectral efficiency (SE) and energy efficiency (EE) respectively. Simulation results show that SCMA performs better than OFDMA and CDMA both in SE and EE.
\end{abstract}


%
\IEEEpeerreviewmaketitle
\section{Introduction}
%
%
%
%

\lettrine{I}{n} the history of mobile communication network development~\cite{fehske2011global}, multiple access technique has evolved from frequency division multiple access (FDMA), time division multiple access (TDMA) and code division multiple access (CDMA) to orthogonal frequency division multiple access (OFDMA)~\cite{wu2014ofdma,wang2012papr,wang2010papr}, where a fundamental cornerstone is the orthogonality of resource block (RB). However, in the next generation of mobile communication system, the demand for massive connection, low latency, high spectral efficiency and energy efficiency has become a vital necessity. Indeed, this scenario may make a turning point from orthogonality to non-orthogonality\cite{wei2017low,wu2018noma}. Sparse code multiple access (SCMA) as a potential non-orthogonal multiple access technique is the key technology of the 5th generation mobile communication systems (5G). It allows multiple users share the same RB and offers 300\% overloading number of user access links~\cite{nikopour2013sparse,wei2019mpa,wei20185g,shen2017scma,liu2019d2d,wei2016lsd,wei2017grantfree}.

SCMA first introduced in \cite{nikopour2013sparse}, is developed from the low density signature (LDS). In SCMA the user codeword is mapped directly into the layer vector, while in LDS, the user codeword is repeated in layer vector element. Therefore SCMA has more diversity gain compared with LDS. SCMA has sparsity in spreading sequences, which allows to use a near optimal message passing algorithm (MPA) algorithm to decode. A systematic approach to optimize the SCMA codebooks has been proposed in \cite{taherzadeh2014scma} based on the design principles of lattice constellations. Futhermore, a resource allocation scheme for SCMA has been proposed in \cite{zhao2015resource}. 

\cite{yu2019maximizing} evaluated the SCMA average spectral efficiency (SE) with adaptive codebooks based on star-QAM signaling constellations, while \cite{wang2017spectral} tested the spectral efficiency improvement for three 5G key performance index. 
\cite{yang2017spectral} compared the spectral-efficiency in SCMA system with three operational scenarios to evaluate their SE performance. Besides, \cite{tong2015enabling} presented 5G radio access technologies for very large number of links which was achieved by SCMA based on optimized sequence design. In addition to analysis and optimization of spectral efficiency studies, SCMA is also compared with the existing techniques~\cite{nikopour2014scma} for the practical performances. It is shown that SCMA outperforms OFDMA in terms of throughput and coverage in practical scenarios. The SCMA area spectral efficiency for cellular network is analyzed via stochastic geometry in \cite{liu2015modeling} and \cite{liu2016performance}, which shows that SCMA is a competitive technique for 5G massive access. 


On the other hand, the SCMA energy efficiency (EE) is also extensively investigated in literatures such as~ \cite{isheden2012framework,ng2012energy,du2014distributed,wu20175g,wu2015powered,wu2016ee}.
An attempt was taken in \cite{zhai2016rate} which maximize the SCMA rate and energy for the wireless powered communication networks. Similarly, \cite{zhang2014sparse} analyzes the energy efficiency for non-orthognal multple access network by providing an analytic framework, which is also used to derive and simulate the SCMA energy efficiency in the uplink scheme. Through simulation and prototype measurement, \cite{dong2016energy} investigated a method to solve the EE maximization problem with sum rate requirements. Since EE maximization problem is a non-convex fractional programming problem \cite{zappone2015energy}, we use symbol transformation and utilize Dinkelbach method to transfer the non-convex problem to a convex problem and derive the global maximum by using an iterative method~\cite{wei2012cfn,wang2011twoway,chen1998sampling,chen2002sampling,li2015caching}.

In this paper, we will investigate the spectral efficiency and energy efficiency of SCMA with an individual rate requirement, which have not been studies before.  Since it is diffcult to derive the analytic solution of the SE and EE optimization problem, we developed iterative methods to solve the SE and EE optimization problem. Since EE maximization is a non-convex problem, the Dinkelbach method is used to transfer the non-convex problem to a convex problem and then solved by iterative method. Simulations show that the SCMA outperforms CDMA and OFDMA both in the terms of SE and EE with QoS. 

\section{System model}
Consider a scenario of single-cell uplink in SCMA networks with $K$ users, $N$ subcarriers and $M$ codebooks, where the base station (BS) and users are equipped with a single antenna. The sets of users and subcarriers are respectively denoted by $\textit{K}$=$\left \{1,2,...,K \right \}$ and $\textit{N}$=$\left \{1,2,...,N \right \}$.
SCMA codebook and subcarriers are the basic resource units \cite{nikopour2013sparse}, \cite{nikopour2014scma}, similar as OFDMA \cite{li2015energy},\cite{song2005cross}. 
Consider overloading access, the user $K$ is greater than the subcarrier $N$, and $K/N$ is called overloading rate. Each user has a codebook and each codeword of size $N$ in the codebook has sparsity. Let $d$ be the non-zero element in a codeword. Then $d\ll N$.

\begin{figure}[H]
	\centering
	\includegraphics[width=3.5in]{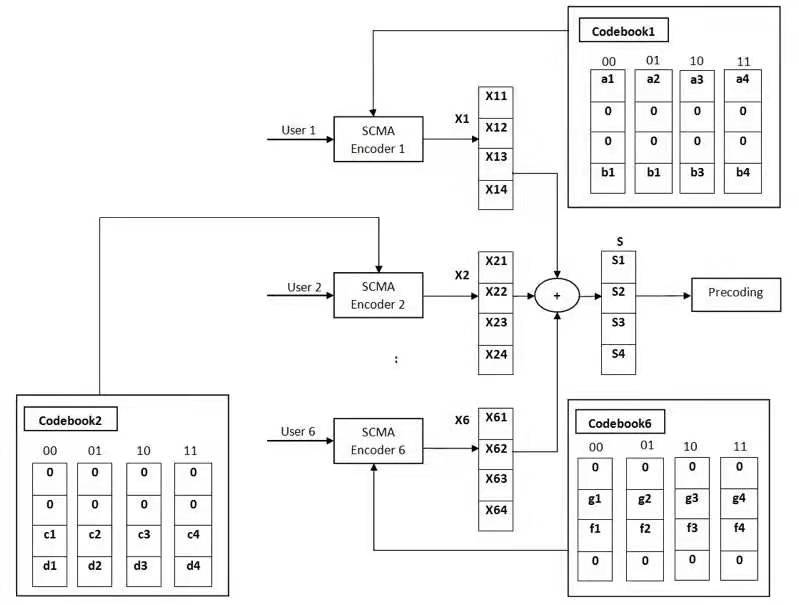}
	\caption{SCMA system network block diagram.}
	\vspace{-0.3em}
	\label{Mapping}
\end{figure}
The SCMA system network diagram with $6$ users and $4$ subcarriers is shown in Fig.~1. Because of the sparsity of SCMA codeword, the receiver can use message passing algorithm (MPA) to detect multi-user. For MPA receiver, codewords allocated to different layers can be regarded as orthogonal resources, so the interference only occurs among users using the same layer. 

The SCMA uplink system, with minimum rate requirement of all users, aims to maximize the total system rate, and establishes an optimal mathematical model for the power allocation and SCMA layer allocation problem. Based on the advocacy of green communication, it is desirable to minimize the transmission power of the system under certain communication guarantees. Therefore, the power allocation scheme is needed. The important content of the research needs to manage the power resources reasonably and effectively according to the specific optimization objectives.

Set up $M$ codebook labels. Set an SCMA codebook and its associated subcarriers can be regarded as a layer of SCMA resources. So there are $N$  SCMA layer resources on each time-frequency resource. 
Therefore, different users obtain multiple access by sharing time-frequency resources on SCMA layer. The codebook size is defined by the length of the codeword and the number of non-zero elements. Assume that the number of SCMA layers in a time-frequency resource block is $m$, and the number of subcarriers is $n$. Indicator variable $c_{nm}$ represents the mapping between the SCMA layer $m$ and the subcarrier $n$. So if layer $m$ occupies subcarrier $n$, then $c_{nm}=1 $ otherwise $0$.

Define $S=\left \{ s_{k,m} \right \}$ as SCMA layer distribution matrix, where $s_{k,m}= 1$ when layer $m$ is assigned to user $k$, otherwise $s_{k,m}= 0$. Define $P={\left \{ p_{k,m} \right \}}$ as power allocation matrix. Then, the total power can be written as

\begin{equation}\label{e1}
P^{tot}=\sum_{k=1}^{K}\sum_{m=1}^{M}s_{k,m}p_{k,m}.
\end{equation}
The power ratio factor assigned to user $k$ using subcarrier $n$ on SCMA layer $m$ is defined as $\alpha_{n,m}$, where $0<\alpha_{n,m}<1$ and $\sum_{n=1}^{N}\alpha_{n,m}=1$. Then, the SNR \cite{bao2016joint}, \cite{cai2017tight} of user $k$ over layer $m$ is 

\begin{equation}\label{e2}
SNR_{k,m}=\frac{\sum_{n=1}^{N}\alpha _{n,m}p_{k,m}h_{k,n}}{\sigma_{k}^{2}},
\end{equation}
where $h_{k,n}$ is the channel state information from user $k$ to base station on subcarrier $n$, $\sigma_{k}^{2}$ represents the noise power to user $k$ with white additive Gaussian noise. Therefore, the achievable rate of user $k$ is

\begin{equation}\label{e3}
R_{k}=\sum_{m=1}^{M}s_{k,m}\log_{2}(1+SNR_{k,m}).
\end{equation}

\section{SCMA Spectral Efficiency with QoS}
In order to maximize the network rate, we take into consideration the constraints of QoS (Quality of Service). Let $w_{k}$ denote the weight factor,  $R_{k}^{req}$ be the minimum data rate requirement of user $k$, and $P^{\max}$ be the maximum transmission power of system. The optimization model can be established as follows.
\begin{equation}\label{e4}
\begin{split}
&\max_{P,S}\sum_{k=1}^{K}w_{k}R_{k},\\
s.t.\,\, &C_{1}: R_{k}\geq R_{k}^{req},\, \forall k,\\
&C_{2}: P^{tot}\leq P^{\max},\\
&C_{3}: \sum_{k=1}^{K}s_{k,m}\leq 1,\, \forall m,\\
&C_{4}: s_{k,m}\in \left \{ 0,1 \right \},\, \forall k,m,\\
&C_{5}: p_{k,m}\geq 0,\,\forall k,m.
\end{split}
\end{equation}
The physical meaning of the optimization problem (\ref{e4}) is to jointly consider SCMA layer allocation and power allocation, by maximizing the network capacity. 
The constraint $C_{1}$ guarantees that the data rate of each user is lower-bounded for fairness. The constraint $C_{2}$ requires that the total transmission power is below the maximum transmission power. The constraint $C_{3}$ and $C_{4}$ jointly ensure that each SCMA layer is allocated to one user at most. Constraint $C_{5}$ is made to ensure that the layer allocated by the user is powered. Note that problem (\ref{e4}) is a mixed integer optimization problem. 


We use the Lagrange dual decomposition method to solve the optimization problem (\ref{e4}). Its partial Lagrange function is
\begin{multline}\label{e7}
L\left ( X,S,\lambda ,\mu  \right )\\
=\sum_{k=1}^{K}\left [ w_{k}R_{k}+\lambda _{k}(R_{k}-R_{k}^{req}) \right ]+\mu (P^{\max}-P^{tot}),
\end{multline}
where $\lambda =\left \{ \lambda _{1},\lambda _{2},\dots,\lambda _{k} \right \}$ and $\mu$ are Lagrange multipliers. Then its dual problem is
\begin{equation}\label{e9}
\min_{\lambda ,\mu }\max_{X,S}L( X,S,\lambda ,\mu), \,\forall \lambda ,\mu \geq 0.
\end{equation}

The optimal $\left \{p_{k,m} \right \}$ can be obtained by finding the partial derivative of (\ref{e9}) and making it equal to zero, i.e.,
\begin{equation}
\frac{\partial L}{\partial p_{k,m}}=0.
\end{equation}
Then 
the optimal power can be calculated as follows.
\begin{equation}\label{e11}
\hat{p}_{k,m}=\left [\frac{w_{k}+\lambda _{k}}{\mu \ln{2}} -\frac{\sigma _{k}^{2}}{\sum_{n=1}^{N}\alpha _{n,m}h_{k,n}} \right ]^{+},
\end{equation}
where $\left [x \right ]^{+}=\max(0,x)$, means that its takes $x$ if $x> 0$ the value is $x$ and $0$ otherwise.
Using the derived optimal power $\hat p_{k,m}$, consider the partial derivative of $L$ with respect to $s_{k,m}$.
\begin{equation}
\frac{\partial L}{\partial s_{k,m}}=H_{k,m},
\end{equation}
where
\begin{multline}\label{e13}
H_{k,m}=(w_{k}+\lambda _{k})\log_{2}\left(1+\frac{\sum_{n=1}^{N}\alpha _{n,m}\hat{p}_{k,m}h_{k,n}}{\sigma _{k}^{2}}\right)\\
-\frac{w_{k}+\lambda_{k}}{\ln{2}}\times \frac{\sum_{n=1}^{N}\alpha_{n,m}\hat{p}_{k,m}}{h_{k,n}} \sigma_{k}^{2}\\
	+\sum_{n=1}^{N}\alpha _{n,m}\hat{p}_{k,m}h_{k,n}-\mu \hat{p}_{k,m}.
\end{multline}
Then $m$ will be assigned to the user $k^{\bullet }$ with the maximum $H_{k,m}$, i,e.,
\begin{equation}\label{e14}
\hat s_{k^{\bullet},m}=1|_{k^{\bullet } ={\arg\max}_k H_{k,m}},\,\forall m.
\end{equation}
Subsequently, the Lagrangian multiplier $\lambda$ and $\mu$ can be updated by the following formulas.
\begin{equation}\label{e15}
\begin{split}
\lambda _{k}(l+1)&=[\lambda _{k}(l+1)-\beta(R_{k}(l)-R_{k}^{req})]^{+},\\
\mu(l+1)&=[\mu(l+1)-\beta(P^{\max}-P^{tot}(l))]^{+},
\end{split}
\end{equation}
where $\beta$ is the iteration step size. Through iterations of (\ref{e11}), (\ref{e14}) and (\ref{e15}), the optimal solution of (\ref{e7}) can be obtained, where one of the metric for the convergence of the iteration is such that ${\max}_k H_{k,m}$ close to $0$. This is summarized in Algorithm~1.
\begin{algorithm}[H]
	\textbf{Initialization} the multipliers $\lambda_k(0)$ and $\mu(0)$, tolerance $\epsilon$;\\
	\textbf{Step 1:} {Update $\hat p_{k,m}$ by (\ref{e11}),}\\
	\textbf{Step 2:} {Update $\hat s_{k,m}$ by (\ref{e13}) and (\ref{e14}),}\\
	\textbf{If} {$\max H_{k,m}>\epsilon$}\\
	\textbf{Then} $l=l+1$, goto Step 1,\\
	\textbf{End If}\\
	{Output $\hat p_{k,m}$, $\hat s_{k,m}$.}
	\caption{SCMA Spectrum Efficiency}
\end{algorithm}

\section{SCMA Energy Efficiency with QoS}
In this section, to maximize the energy efficiency of the system, the objective function is established. Based on the quasi-convex optimization theory, the objective function is analyzed, and a joint power and SCMA layer assignment algorithm is proposed, which improves the network energy efficiency while satisfying all individual users' QoS requirements.
%
The total power consumption of the SCMA system is
\begin{equation}
P=\varepsilon _{0}P^{tot}+P_{0},
\end{equation}
where the coefficient $\varepsilon_0$ is the power amplifier factor, and $P_{0}$ is the circuit power. According to (\ref{e1}), (\ref{e2}) and (\ref{e3}), the system energy efficiency (EE) is defined by
\begin{equation}
\eta_{EE}=\frac{R}{P}=\frac{\sum_{k=1}^{K}R_{k}}{\varepsilon _{0}\sum_{k=1}^{K}\sum_{m=1}^{M}s_{k,m}p_{k,m}+P_{0}}.
\end{equation}

Based on the system model, the energy efficiency of power-constrained single cell multi-user networks is planned by joint power allocation and codebook allocation, and formulated as follows.
\begin{equation}\label{p3}
\begin{split}
&\max_{P,S}\frac{\sum_{k=1}^{K}R_{k}}{\varepsilon _{0}\sum_{k=1}^{K}\sum_{m=1}^{M}s_{k,m}p_{k,m}+P_{0}},\\
s.t.\,\, &C_{1}:R_{k}\geq R_{k}^{req},\,\forall k,\\
&C_{2}: P^{tot}\leq P^{max},\\
&C_{3}: \sum_{k=1}^{K}s_{k,m}\leq 1,\forall m,\\
&C_{4}:s_{k,m}\in \left \{ 0,1 \right \},\,\forall k,m,\\
&C_{5}:p_{k,m}\geq 0,\,\forall k,m.
\end{split}
\end{equation}
%
In the optimization problem (\ref{p3}), the constraint $C_{1}$ guarantees that individual user's rate meets its minimum rate requirement satisfying the QoS, $C_{2}$ requires that the total transmission power is not greater than its maximum transmission power, $C_{3}$ and $C_{4}$ ensures that each user can allocate up to one layer of SCMA resources, and lastly $C_{5}$ ensures the power allocated by users to SCMA layer is non-negative. 

The optimization problem (\ref{p3}) is a fractional programming problem with combinatorial properties and belongs to non-convex optimization problem, it is hard to directly solve this problem. In order to facilitate operation, define $X=\{x_{k,m}|x_{k,m}=s_{k,m}p_{k,m} \}$. Therefore, the problem (\ref{p3}) can be rewritten as follows.

\begin{equation}\label{p4}
\small
\begin{split}
&\max_{X,S}\frac{\sum_{k=1}^{K}\sum_{m=1}^{M}s_{k,m}\log_{2}(1+SNR_{k,m})}{\varepsilon _{0}\sum_{k=1}^{K}\sum_{m=1}^{M}x_{k,m}+P_{0}},\\
s.t.\,\, &C_{1}:\sum_{m=1}^{M}s_{k,m}\log_{2}\left(1+\frac{\sum_{n=1}^{N}\alpha _{n,m}x_{k,m}h_{k,n}}{\sigma _{k}^{2}s_{k,m}}\right)\geq R_{k}^{req},\\
&C_{2}:\sum_{k=1}^{K}\sum_{m=1}^{M}x_{k,m}\leq P^{max},\\
&C_{3}:\sum_{k=1}^{K}s_{k,m}\leq 1,\,\forall m,\\
&C_{4}:0\leq x_{k,m}\leq 1,\,\forall k, m,\\
&C_{5}:x_{k,m}\geq 0,\,\forall k, m.
\end{split}
\end{equation}
At this stage, the optimization problem (\ref{p4}) is still a non-convex optimization problem, which needs further transformation. Assuming that the objective function of the optimization problem (\ref{p4}) is $q$, and write
\begin{multline}
F(q,s_{k,m},x_{k,m})\overset\triangle =\sum_{k=1}^{K}\sum_{m=1}^{M}s_{k,m}\log_{2}\left(1+SNR_{k,m}\right)\\
-q\left(\varepsilon _{0}\sum_{k=1}^{K}\sum_{m=1}^{M}x_{k,m}+P_{0}\right).
\end{multline}
If $F(q,s_{k,m},x_{k,m})=0$, the optimal solution $(\hat{s}_{k,m})$ and $(\hat{x}_{k,m})$ of the optimization problem (\ref{p4}) in term of $q$ is obtained, and $q$ is the optimal EE.
%
Therefore, the original optimization problem can be further transformed into
\begin{equation}\label{p5}
\small
\begin{split}
&\max_{X,S}F(q,s_{k,m},x_{k,m}),\\
s.t.\,\, & C_1, C_2, C_3, C_4, C5 \text~in~problem~({\ref{p4}}).
\end{split}
\end{equation}

At this stage, the optimization problem (\ref{p5}) is a convex optimization problem, and Lagrange function is

\begin{multline}\label{e22}
L(X,S,\lambda ,\mu )=\sum_{k=1}^{K}R_{k}-q\left(\varepsilon_{0}\sum_{k=1}^{K}\sum_{m=1}^{M}x_{k,m}+P_{0}\right)\\
+\sum_{k=1}^{K}\lambda _{k}(R_{k}-R_{k}^{req})+\mu (P^{\max}-P^{tot}),\\
=\sum_{k=1}^{K}R_{k}(1+\lambda _{k})-\sum_{k=1}^{K}\lambda _{k}R_{k}^{req}\\
-q\varepsilon _{0}\sum_{k=1}^{K}\sum_{m=1}^{M}x_{k,m}-\mu P^{tot}-qP_{0}+\mu P^{\max},
\end{multline}
where $\lambda= \left \{ \lambda _{1},\lambda _{2},..., \lambda _{k} \right \}$ and $\mu$ are Lagrange multipliers.
Its dual problems is
\begin{equation}
\min_{\lambda ,\mu }\max_{X,S}L(X,S,\lambda,\mu), \,\lambda ,\mu \geq 0.
\end{equation}
The optimal $x_{k,m}$ can be obtained by finding the partial derivative of (\ref{e22}) and making it equal to $0$, i.e.,
\begin{equation}
\frac{\partial L(X,S,\lambda,\mu )}{\partial x_{k,m}}=0.
\end{equation}
This lead to
\begin{equation}
x_{k,m}=\frac{(1+\lambda _{k})s_{k,m}}{(\mu +q\varepsilon _{0})\ln_{2}}-\frac{\sigma_{k}^{2}s_{k,m}}{\sum_{n=1}^{N}\alpha _{n,m}h_{k,n}}.
\end{equation}
Hence the optimal power can be written as
\begin{equation}\label{e26}
\small
\hat{p}_{k,m}=\frac{x_{k,m}}{s_{k,m}}=\left [ \frac{(1+\lambda _{k})}{(\mu+q\varepsilon _{0} )\ln{2}} -\frac{\sigma _{k}^{2}}{\sum_{n=1}^{N}\alpha _{n,m}h_{k,n}}\right ]^{+}.
\end{equation}
The partial derivative $L$ with respect to $s_{k,m}$ is as follows.
\begin{equation}
\frac{\partial L(X,S,\lambda,\mu )}{\partial s_{k,m}}=H_{k,m},
\end{equation}
where
\begin{multline}\label{e28}
\small
H_{k,m}=(1+\lambda _{k})\log_{2}\left(1+\frac{\sum_{n=1}^{N}\alpha _{n,m}\hat{p}_{k,m}h_{k,n}}{\sigma _{k}^{2}}\right)\\
-\frac{1+\lambda _{k}}{(\mu +q\varepsilon _{0})\ln{2}}\times\frac{\sum_{n=1}^{N}\alpha _{n,m}\hat{p}_{k,m}h_{k,n}}{\sigma _{k}^{2}+\sum_{n=1}^{N}\alpha _{n,m}\hat{p}_{k,m}h_{k,n}}-\mu \hat{p}_{k,m}.
\end{multline}
The codebook layer $m$ should be allocated to have the maximum $H_{k,m}$:
\begin{equation}\label{e29}
\hat{s}_{k^{\bullet},m}=1|_{k^{\bullet}=\arg\max_{k} H_{k,m}}, \,\forall m
\end{equation}

According to the sub-gradient algorithm, the Lagrangian multipliers $\lambda_k$ and $\mu$ can be updated as follows.
\begin{equation}\label{e30}
\begin{split}
\lambda_{k}(l+1)&=[\lambda _{k}(l+1)-\beta(R_{k}(l)-R_{k}^{req})]^{+},\\
\mu(l+1)&=[\mu(l+1)-\beta(P^{\max}-P^{tot}(l))]^{+},
\end{split}
\end{equation}
where $\beta$ is the iteration step.  Through iterations of (\ref{e26}), (\ref{e29}) and (\ref{e30}), the optimal solution of (\ref{e22}) can be obtained, where the metrics for the convergence of the iteration are such that $\max_{X,S} F(q,s_{k,m},x_{k,m})$ and ${\max}_k H_{k,m}$ close to $0$. This is summarized in Algorithm~2.


\begin{algorithm}[H]
	\textbf{Initialization}: The multipliers $\lambda_k(0)$ and $\mu(0)$, the energy efficiency $q$, tolerance $\epsilon>0$;\\
	\textbf{Step 1:} {Update the power allocation $\hat p_{k,m}$ by (\ref{e26});}\\
	\textbf{Step 2:} {Update the assignment index $\hat s_{k,m}$ by (\ref{e28}) and (\ref{e29});}\\
	\textbf{Step 3:} {Update $q$, update $\lambda_k$ and $\mu$ by (\ref{e30});}\\
	\textbf{If} {$\max F(q,p_{k,m},x_{k,m})>\epsilon$,\\
		\textbf{Then} $l=l+1$; goto Step 1,}\\
	\textbf{End If}\\
	Output the maximum EE $q$, $s_{k,m}$  and $p_{k,m}$.
	\caption{SCMA Energy Efficiency}
\end{algorithm}

\section{Simulation results}
In this simulation section, the Algorithm~1 and the Algorithm~2 are simulated and compared. The simulation parameters are given in the Table~1. The following Fig. \ref{SpectralEfficiencyComparaison} and Fig. \ref{EnergyEfficiencyComparaison} respectively show a comparison of the spectral performance and energy efficiency for SCMA, CDMA, and OFDMA. It is found that SCMA perfoms better than CDMA and OFDMA both in SE and EE.
\begin{table}[H]
	 \centering
	\caption{Simulation parameters.}
	\begin{tabular}{|c|c|}
		\hline
		\textbf{Parameter}                 & \textbf{Value}   \\ \hline
		Number of subcarriers    & 4       \\ \hline
		Number of codebooks      & 6       \\ \hline
		Circuit power consumption & 1w      \\ \hline
		R                         & 500m    \\ \hline
		h                         & 30m     \\ \hline
		Subcarrier bandwidth      & 156kHz  \\ \hline
		SCMA layer               & 12      \\ \hline
		Noise power               & -112dBm \\ \hline
		Iteration times           & 100     \\ \hline
		Weight factor             & 1       \\ \hline
		$P_{0}$                     & 1w      \\ \hline
		$R_{k}^{req}$               & 120Kbps \\ \hline
		$P^{max}$                   & 100w    \\ \hline
		Power amplifier factor    & 1/0.37  \\ \hline
	\end{tabular}
\end{table}

Fig. 2 shows that SCMA spectral efficiency outperforms both OFDMA and CDMA. Meanwhile, OFDMA had a better spactral efficiency performance comparing to CDMA at roughly 50 BSs, where CDMA spectral efficiency remains stable; hence, OFDMA supports higher modulation and coding leading to a better spectral efficiency, enhancement of reachability and provides a significant improvement in spectral usage comparing to CDMA tehcnology performance.

Fig. 3 illustrates a significant decrease of OFDMA energy efficiency when the number of users increases due to the high level consumption of the power; while SCMA proves a better energy efficiency achievement. 

\begin{figure}[H]
	\centering
	\includegraphics[width=3.5in]{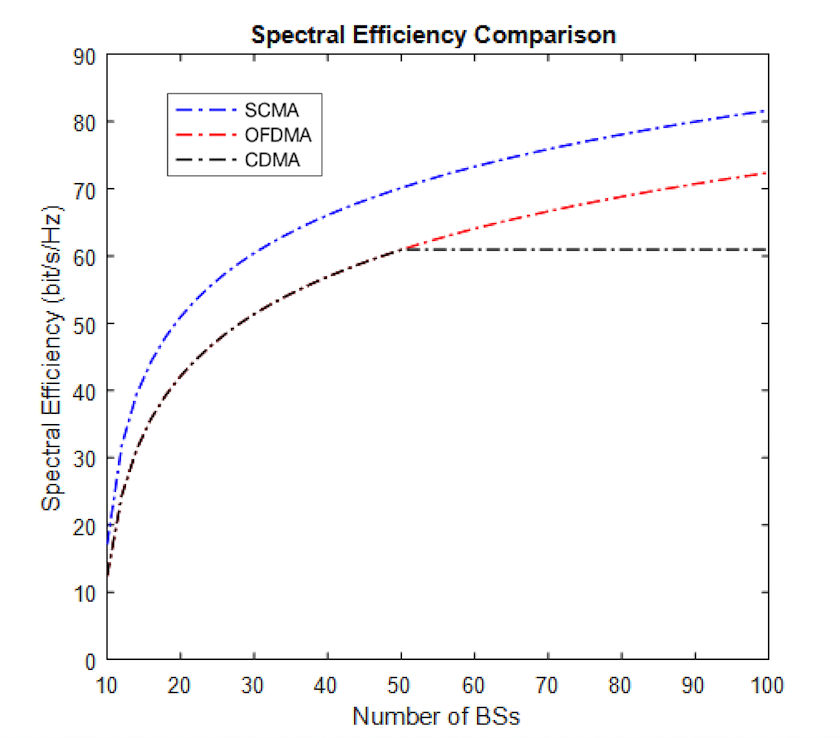}
	\caption{The SE performance for SCMA, OFDMA and CDMA.}
	\vspace{-0.3em}
	\label{SpectralEfficiencyComparaison}
\end{figure}

\begin{figure}[H]
	\centering
	\includegraphics[width=3.5in]{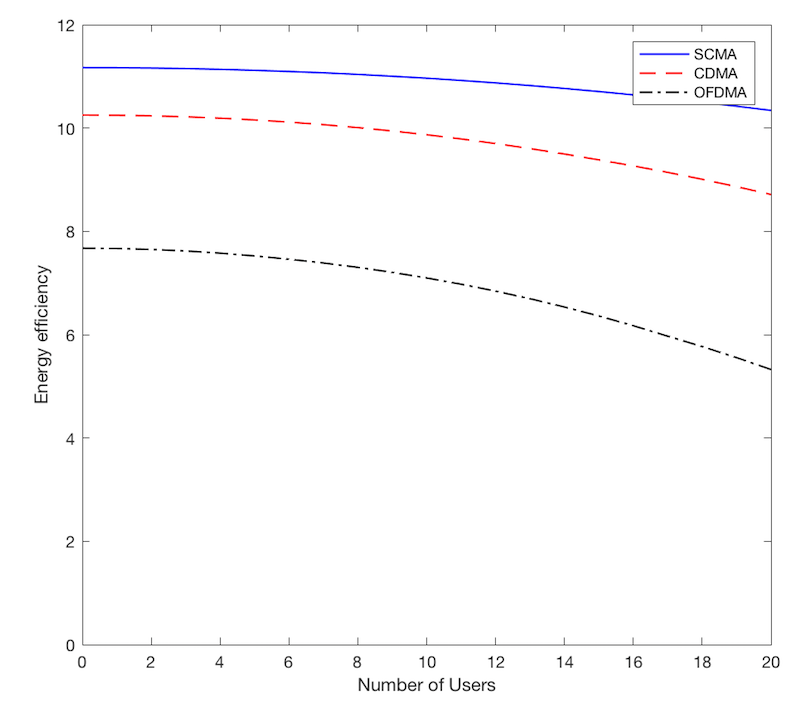}
	\caption{The EE performance for SCMA, OFDMA and CDMA.}
	\vspace{-0.3em}
	\label{EnergyEfficiencyComparaison}
\end{figure}


\section{Conclusion}
This paper 
maximizes the spectral efficiency and the energy efficiency for SCMA network with the individual user's rate requirements. The formulated optimization problem is fractional programming non-convex problem. Using symbol transformation and Dinkelbach method, the original problem is transformed into convex problems, and solved by Lagrange dual decomposition method. The proposed algorithms are simulated to compare with CDMA and OFDMA. It is found that the SCMA performs better than CDMA and OFDMA both in SE and EE, since the proposed method has shown a better network mapping matrix based on power allocation and codebook assignment. In perspective, the continuity of this paper will discuss the impact of the number of subcarriers and SCMA layers in the system as well as its feasibility and expertise.


\section*{Acknowledgment}
This paper is supported in part by National Key Project 2018YFB1801102, and in part by NSFC 61671294.

\bibliographystyle{IEEEtran}
\bibliography{RefBibTex}
\end{document}